\documentclass[twocolumn,tighten]{aastex63}
\accepted{02/24/2021}
\received{12/18/2020}
\submitjournal{AAS journals}
\usepackage{graphicx}
\usepackage{natbib}
\usepackage{latexsym}
\usepackage{amssymb}
\usepackage{longtable}
\usepackage{amsmath}
\usepackage{url}
\def\msun{\ifmmode {\rm\,M_\odot}\else ${\rm\,M_\odot}$\fi}
\def\Msun{\ifmmode {\rm\,\it{M_\odot}}\else ${\rm\,M_\odot}$\fi}
\def\lsun{\ifmmode {\rm\,L_\odot}\else ${\rm\,L_\odot}$\fi}
\def\Lsun{\ifmmode {\rm\,\it{L_\odot}}\else ${\rm\,L_\odot}$\fi}
\def\rsun{\ifmmode {\rm\,R_\odot}\else ${\rm\,R_\odot}$\fi}
\def\Rsun{\ifmmode {\rm\,\it{R_\odot}}\else ${\rm\,R_\odot}$\fi}
\def\Tsun{\ifmmode {\rm\,T_\odot}\else ${\rm\,T_\odot}$\fi}
\def\arcsec{\ifmmode {^{\prime\prime}}\else $^{\prime\prime}$\fi}
\def\asec{\ifmmode {^{\prime\prime}}\else $^{\prime\prime}$\fi}
\def\arcmin{\ifmmode {^{\prime}}\else $^{\prime}$\fi}
\def\amin{\ifmmode {^{\prime}}\else $^{\prime}$\fi}
\def\simlt{\mathrel{\spose{\lower 3pt\hbox{$\mathchar"218$}}
     \raise 2.0pt\hbox{$\mathchar"13C$}}}
\def\simgt{\mathrel{\spose{\lower 3pt\hbox{$\mathchar"218$}}
\     \raise 2.0pt\hbox{$\mathchar"13E$}}}

\def\htwo{H$_2$}
\def\hst{\textit{HST}}

\begin{document}

\author{P. Wilson Cauley}
\affiliation{Laboratory for Atmospheric and Space Physics, University of Colorado Boulder, 600 UCB, Boulder, CO 80303}

\author{Kevin France}
\affiliation{Laboratory for Atmospheric and Space Physics, University of Colorado Boulder, 600 UCB Boulder, CO 80303}

\author{Gregory J. Herzceg}
\affiliation{Kavli Institute for Astronomy and Astrophysics, Peking University, Yiheyuan Lu 5, Haidian Qu, 100871 Beijing, People’s Republic of China}

\author{Christopher M. Johns-Krull}
\affiliation{Physics \& Astronomy Department, Rice University, 6100 Main Street, Houston, TX 77005}

\correspondingauthor{P. Wilson Cauley}
\email{pwcauley@gmail.com}

\title{A CO-to-H$_2$ ratio of $\approx 10^{-5}$ towards the Herbig Ae star HK Ori}

\begin{abstract} 

Measurements of gas mass in protoplanetary gas disks form the basis for estimating the conditions of 
planet formation. Among the most important constraints derived from disk diagnostics are the 
abundances of gas-phase species critical for understanding disk chemistry.
Towards this end, we present direct line-of-sight measurements of H$_{2}$ and CO, 
employing UV absorption spectroscopy from $HST$-COS to characterize disk composition,
molecular excitation temperatures, and spatial distribution in the circumstellar material
around the Herbig Ae stars HK Ori and T Ori.  We observe strong CO (N(CO) = 10$^{15.5}$ cm$^{-2}$; 
T$_{rot}$(CO) = 19 K) and H$_{2}$ (N(H$_{2}$) = 10$^{20.34}$ cm$^{-2}$; T$_{rot}$(H$_{2}$) = 141 K) absorption
towards HK Ori with a CO/H$_{2}$ ratio ($\equiv$ N(CO)/N(H$_{2}$)) = 1.3$^{+1.6}_{-0.7}$~$\times$~10$^{-5}$.  
These measurements place direct empirical constraints on the CO-to-H$_{2}$ conversion factor 
in the disk around a Herbig Ae star for the first time, although there is uncertainty
concerning the exact viewing geometry of the disk. The spectra of T Ori show CO 
(N(CO) = 10$^{14.9}$ cm$^{-2}$; T$_{rot}$(CO) = 124 K) absorption. Interestingly, we do not 
detect any H$_{2}$ absorption towards this star (N(H$_{2}$) $<$ 10$^{15.9}$ cm$^{-2}$).  
We discuss a potential scenario for the detection of CO without H$_{2}$, which deserves
further investigation. The low abundance ratio measured around HK Ori suggests significant 
depletion of CO in the circumstellar gas, which conforms with the handful of other recent CO 
abundance measurements in protoplanetary disks.  

\end{abstract}


\section{INTRODUCTION}
\label{sec:intro}

In the inner disks ($r$~$<$~10 AU) around young stars, molecular gas emission and absorption
provide our best means of inferring the physical conditions at the radii where gas
giant and rocky planet cores are forming and accreting their nascent
atmospheres.  The abundances of gas-phase species at the location of formation
and accretion are expected to affect the chemical abundances of exoplanet
atmospheres \citep{oberg11}. Recent surveys of molecular emission from
Classical T Tauri Stars (CTTSs) and Herbig Ae stars have provided new
constraints on the radial distribution, temperature, and composition of
planet-forming disks. Surveys of mid-IR emission from CO 
\citep{salyk09,banzatti15}, H$_{2}$O and organic molecules
\citep{pontoppidan10,carr11}, and spectrally/spatially-resolved
near-IR observations \citep{pontoppidan11,brittain15} have
placed constraints on the CO and H$_{2}$O inventory at planet-forming radii and
the evolution of the inner molecular gas disk radius \citep{antonellini20}. 
All of these studies require an accurate knowledge of the absolute abundances 
to convert these measurements into local gas masses.

An emerging consensus from ALMA observations of protoplanetary disks is that on large
scales ($\sim 30-100$ AU), CO is heavily depleted relative to \htwo\ \citep[e.g.,][]{miotello17,long17}
This depletion process may be fairly rapid, occurring within the first $\approx 1$ Myrs
of a proto\textit{stellar} disk's lifetime \citep{zhang20}. 
When interpreted with physical-chemical models of disks that include CO freeze-out and
photodissociation \citep{miotello16}, the $^{13}$CO line fluxes are much
weaker than expected from dust emission and the few available detections of HD
\citep{bergin13,mcclure16}, implying that something other than freeze-out is
depleting the CO. The most plausible explanation is that complex
C-bearing molecules also freeze-out in the disk mid-plane, although chemical
reprocessing of CO in the cold disk mid-plane is also possible for 
CTTS disks which are older than $\approx 1$ Myr \citep{bosman18}.
Once frozen out in a low-viscosity region, the ices will remain frozen out, with a time-dependence
that will gradually deplete the disk of C and therefore CO \citep{kama16,schwarz16}. 

Our best opportunity to converge on a three-dimensional view of the
planet-forming regions around young stars, both inside and outside 10 AU, is to
assemble a broad spectrum of relevant disk tracers and diagnostics
\citep{sicilia16}. Among the most important items in this toolkit
of disk diagnostics are measurements of the abundances and the physical
conditions of the most important gas-phase species. Molecular absorption 
spectroscopy, where a pencil beam through the circumstellar disk is observed 
against the host star, provides a one-dimensional slice through the disk and 
allows the column density, temperature, and radial velocity of the absorbing 
gas to be quantified. This has historically been one of the most powerful 
techniques for diagnosing the composition and temperature structure of the ISM, 
and our approach in this work is to apply ultraviolet absorption spectroscopy 
techniques to better understand planet-forming environments around young stars.  

UV absorption line spectroscopy has been used to measure CO around a number of 
CTTSs \citep{mcjunkin13} and resulted in the first circumstellar CO/\htwo\ ratio for a pre-main sequence
star \citep{france14}. UV spectroscopy cannot probe the disk mid-plane, 
but current models predict that carbon should rapidly get sequestered
in solids in the disk mid-plane at all radii and not return to the gas phase.
In the inner disk ($r \lesssim 4$ AU) the CO and H$_2$ in the warm molecular
layer accounts for the vast majority of the carbon and hydrogen respectively 
\citep{adamkovics14}, which suggests that CO/H$_2$ measured through
this region of the disk is a good approximation of the true CO/H$_2$
value.

The focus of this work is the circumstellar gas around the Herbig Ae stars T Ori 
and HK Ori. Both T Ori and HK Ori are well studied pre-main 
sequence systems. T Ori is a UX Orionis (UXOR) variable star
which hosts a self-shadowed circumstellar disk \citep{hillenbrand92,bertelsen16} 
and has an estimated accretion rate of $\text{log}(\dot{M}) = -6.58 \pm 0.40 \Msun\ \text{yr}^{-1}$
\citep[][however, see \autoref{sec:disc}]{mendigutia11}. The leading explanation 
for the UXOR phenomenon is the occultation of the central star by dust in 
the warped or puffed up inner disk rim
\citep{dullemond03,kreplin16}. This requires that the disk be viewed nearly 
edge-on ($i \gtrsim 70$). Although no estimates of T Ori's disk inclination angle exist, 
its nature as a UXOR and the SED evidence of a self-shadowed disk suggests that its 
disk is viewed at a large inclination angle. It also has a stellar rotational velocity 
of $v$sin$i \approx 150 - 175$ km s$^{-1}$ \citep{mora01,alecian13,cauley15} implying 
that the star is viewed equatorially rather than pole-on.

HK Ori is also actively accreting from its circumstellar disk \citep{hillenbrand92,mendigutia11}
with an accretion rate of $\text{log}(\dot{M}) = –5.24 \pm 0.12 \Msun\ \text{yr}^{-1}$
and has a K4 binary companion at $\approx 150$ AU, or $\approx 0\overset{\arcsec}{.}3$
\citep{leinert97,baines04,smith05,wheelwright10}.
There are no published constraints on the inclination angle of HK Ori's circumstellar
disk, although \citet{blondel06} note that the low equivalent width of the \ion{Fe}{2} 2400 \AA\
absorption line suggests that we are viewing the star closer to pole-on since
an edge-on orientation would produce a large \ion{Fe}{2} equivalent width.
\citet{cauley15} derived a projected rotational velocity of $v$sin$i = 20$ km
s$^{-1}$ for HK Ori which is also indicative of a low inclination angle for
the system. However, the spectrum of HK Ori is contaminated by a plethora of
emission lines and there are few clean photospheric absorption lines which can
be used to fit spectral templates. Thus the value from \citet{cauley15} should
be viewed with caution. In addition, HK Ori shows significant spectroscopic
and photometric variability \citep{eiroa02,baines04,mendigutia11a} similar to the UXOR
phenomenon although with smaller amplitude. Finally, HK Ori on average shows
a double-peaked H$\alpha$ line profile with a strong central absorption feature 
\citep{reipurth96,mendigutia11a,cauley15}. This profile shape was shown to be 
qualitatively reproduced by accreting HAe stars viewed at high ($\gtrsim 70^\circ$) 
inclination \citep{muzerolle04}. The H$\alpha$ line profile variability also
agrees well with the UXOR obscuration scenario tested by \citet{muzerolle04} for
UX Ori. Overall, the evidence for a disk with a high inclination angle (i.e., edge-on disk)
is stronger than for a low inclination angle.

In this paper we present new and archival \textit{Hubble Space Telescope} FUV data
for T Ori and HK Ori. We use the UV spectra to measure CO and 
\htwo\ column densities via pencil beam absorption through the circumstellar 
material and then compare the derived column
densities to place constraints on the CO/\htwo\ ratios in the gas. \autoref{sec:obs} describes
the UV spectroscopic data sets and \autoref{sec:models} describes the spectral synthesis
modeling used to derive molecular parameters. \autoref{sec:disc} places these results in
context of current disk and interstellar sight-line studies and a brief summary 
of our findings is presented in \autoref{sec:conc}.

\section{OBSERVATIONS AND DATA REDUCTION}
\label{sec:obs}

The data were collected in 2012 and 2019 as part of the HST Guest Observing
Programs 12996 (P.I. Johns-Krull) and 15070 (P.I. France).
Both HK Ori and T Ori were observed with two different Cosmic Origins
Spectrograph (COS) settings in each program. The COS G130M $\lambda 1222$
setting from Program 15070 was selected to cover the wavelength range 1090 -
1360 \AA\, which includes two H$_2$ $v=0$ Lyman band transitions \citep{abgrall89}. We
note that the G130M $\lambda 1291$ setting from Program 12996 did not reach far
enough into the blue to cover the relevant H$_2$ lines. Thus the H$_2$ analysis
utilizes only the Program 15070 data. The G130M observations have a resolving
power of $R \approx 16,000$, or a velocity resolution of $\Delta v = 19$ km
s$^{-1}$. HK Ori was observed for a total of $t = 7758$ seconds in this mode,
while T Ori was observed for $t = 10089$ seconds.

The COS G160M $\lambda 1577$ setting covers the wavelength range $1400 - 1750$
\AA\ and contains the CO Fourth Positive \textit{A-X} system transitions from
the $0 \leq v' \leq 4$ vibrational states. The G160M observations have a
resolving power of $R \approx 18,000$, or a velocity resolution of $\Delta v =
17$ km s$^{-1}$. HK Ori was observed for a total of $t = 1376$ seconds and
T Ori was observed for $t = 1696$ seconds. All COS exposures were reduced
with the standard \texttt{calCOS} pipeline.

Quality G160M spectra are available for T Ori from both 2012 and 2019 through
both programs. The 2019 spectrum taken through Program 15070 has approximately 
double the amount of flux compared with the 2012 spectrum and the CO absorption 
is weaker by a factor of $\approx 2$. Interestingly, the flux ratio is 
roughly constant at $\approx 2$ between the two epochs for wavelengths $\gtrsim 1300$ 
\AA\ where the photosphere dominates but drops to $\approx 1$ for wavelengths $\lesssim 
1300$ \AA\ where the accretion continuum likely dominates. 


We have opted to use the 2012 spectrum in our analysis because of the more
prominent CO feature. Although our choice to use the 2019 spectrum biases the CO analysis
because of the stronger absorption seen in that epoch, the lack of H$_2$ in T Ori's
spectrum (see \autoref{sec:h2models}) prohibits any useful constraints on
CO/H$_2$ in the system. Thus the bias introduced does not influence any of our
main conclusions concerning the circumstellar material around T Ori. We 
provide a more detailed discussion of T Ori's FUV flux variability and how it
might be related to the variable CO absorption and lack of H$_2$ absorption 
in \autoref{sec:tori_h2}.


\begin{figure*}[htbp]
   \centering
   \includegraphics[scale=.70,clip,trim=0mm 15mm 5mm 25mm,angle=0]{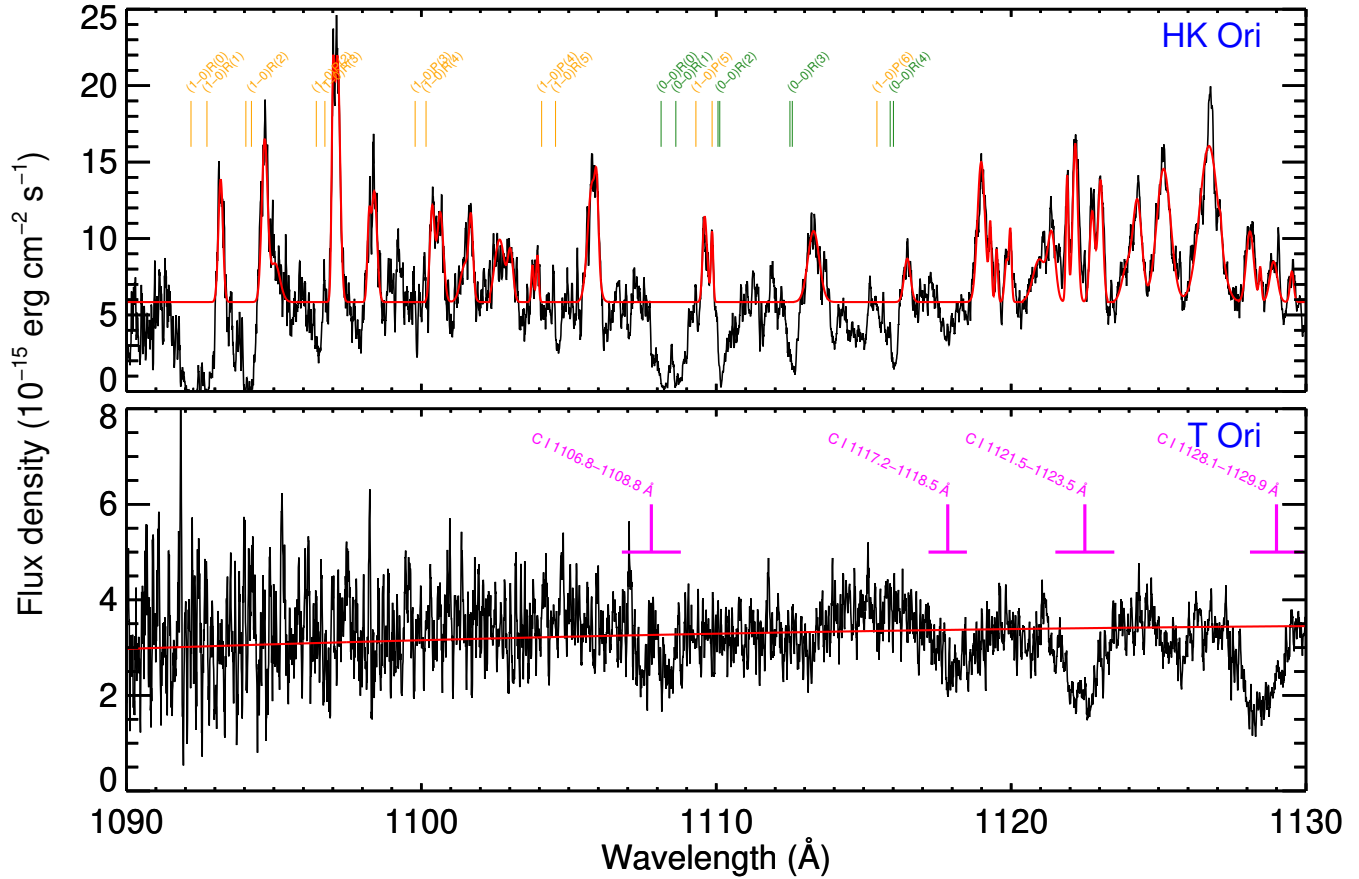}

   \figcaption{Extracted G130M spectra in the surrounding region of the H$_2$
Lyman band. The observed spectra are shown in black and the continuum fits in
red. Transitions to $v^{`} = 0$ are marked in green and transitions to $v^{`} = 1$
are marked in orange. Transition labels for (1-0)P(1) at 1094.05 \AA, (1-0)R(6) at 1109.86
\AA, (0-0)P(1) at 1110.06 \AA, (0-0)P(2) at 1112.50 \AA, and (0-0)P(3) at 1115.90 \AA\ 
are omitted for clarity. Clear absorption is visible in the HK Ori spectrum while 
the T Ori spectrum shows no sign of absorption by H$_2$. The absorption features in the T Ori
spectrum are \ion{C}{1} lines, most of which are likely accretion-related.
\label{fig:h2specs}}

\end{figure*}

\begin{figure*}[htbp]
   \centering
   \includegraphics[scale=.70,clip,trim=0mm 15mm 5mm 25mm,angle=0]{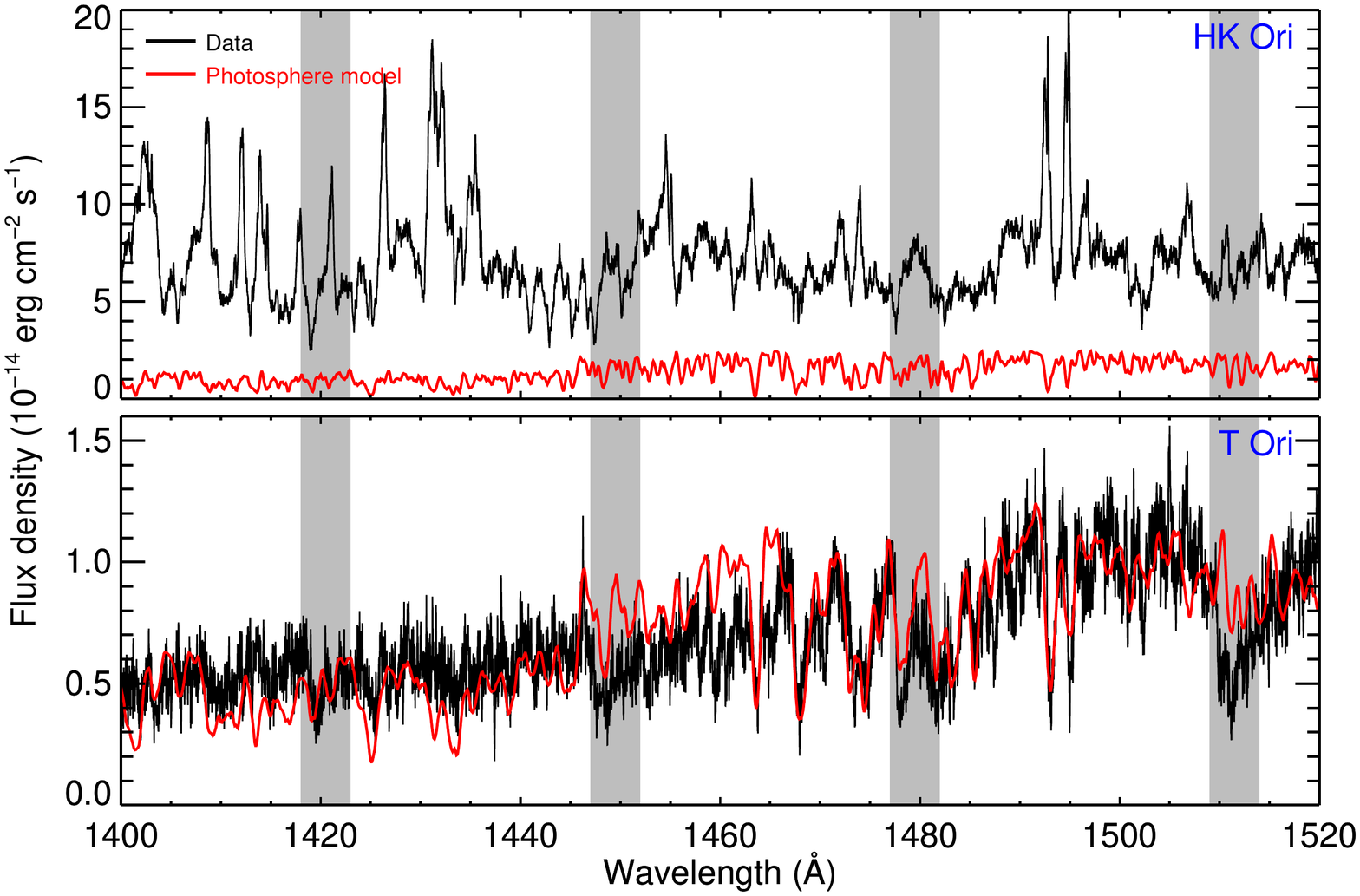}

   \figcaption{Extracted G160 M spectra in the surrounding region of the CO
Fourth Positive system. The stellar spectra are shown in black and
scaled photosphere models from PHOENIX are shown in red. The
photospheric models are used as a guide to which features are stellar versus
circumstellar. The wavelength segments containing the CO $A - X$ bands
are shaded in gray. \label{fig:cospecs}}

\end{figure*}

\autoref{fig:h2specs} show the final co-added spectra surrounding the H$_2$
Lyman band transitions of interest \citep{abgrall89} and \autoref{fig:cospecs} shows the final
co-added spectra bracketing the CO $A - X$ band transitions. The spectra are
smoothed with a 5-pixel boxcar. The UV spectra of many Herbig Ae/Be stars are
littered with emission lines \citep[e.g.,][]{grady96,bouret98,cauley16}, which
are most clearly seen in HK Ori's spectrum. In order to isolate the H$_2$ and CO
absorption, we manually fit a high order spline, combined with a first or
second degree polynomial, to the contaminating emission lines to estimate the
``continuum'' emission. The continuum fits in the $1090 - 1130$ \AA\ region are
plotted as red lines in \autoref{fig:h2specs}. 

For the $1400 - 1520$ \AA\ region containing the CO $A - X$ absorption bands,
we compared the observed spectra to PHOENIX stellar photosphere models
\citep{husser13} to aid in identifying which features are stellar versus
circumstellar. Scaled and reddened PHOENIX models are over-plotted in red in
\autoref{fig:cospecs}. Note that HK Ori is a binary system with an A-type
primary and G-type T Tauri secondary with a separation of $\approx 150$ AU
\citep{leinert97,smith05}. We only show the A-type photosphere model in 
\autoref{fig:cospecs} due to the much weaker flux of
the G-type photosphere in the FUV. It is clear that the photosphere of 
HK Ori is dwarfed by the emission lines so the photosphere is 
relatively unimportant in this case. For T Ori, however, the spectrum 
from $1400 - 1520$ \AA\ is almost entirely due
to the star's photosphere with little, if any, excess accretion emission
present. Thus we closely compared the photosphere model to the observed spectrum
during the spline continuum fitting for T Ori so as to minimize the inclusion
of stellar features in the final normalized CO spectrum. The excess CO
absorption is fairly clear in most of the bandheads in \autoref{fig:cospecs}.

\section{MODELING THE H$_2$ and CO ABSORPTION}
\label{sec:models}

Calculating the CO/H$_2$ ratio in the warm molecular layer of a disk requires a
measurement of the co-spatial CO and H$_2$ densities. To accomplish this we
model the column densities of H$_2$ and CO independently. The temperature of
each gas is also derived independently. For the CO models, the rotational
temperature is a model parameter; the H$_2$ rotational temperature is derived
from the best-fit column densities in a separate fitting procedure. All model
spectra are convolved with the COS linespread-function (LSF) \citep{kriss11},
which varies as a function of wavelength. 

\subsection{H$_2$ models}
\label{sec:h2models} 

The $v^{\prime\prime}=0,J^{\prime\prime}$ H$_2$ absorption lines are fit using
the \textit{h2ools} suite of optical depth templates from \citet{mccandliss03}, which
contain all of the Lyman and Werner transitions from the ground vibrational
level of the electronic ground state \citep{abgrall89}. Each template spectrum is calculated for
an integer value of the Doppler parameter $b$, where 
$b = (2 k T_\text{rot}/m_\text{H2}+v_\text{turb}^2)^{1/2}$, and assuming a column density of
$N = 10^{21}$ cm$^{-2}$. The free parameters in the fit are the
line-of-sight radial velocity of the absorbing gas $v_\text{rad}$, the coverage
fraction of the stellar disk by the gas $f_\text{cov}$, the Doppler broadening
parameter $b$, and the column densities of the rotational transitions
$J^{\prime\prime} = 0 - 8$. We note that although the velocity resolution
of the observations is $\approx 17$ km s$^{-1}$, the value of $b$ can be constrained
at much higher precision since it has a strong effect on the optical
depth of the unsaturated H$_2$ transitions. Because the templates from \citet{mccandliss03}
are optical depth vectors, and the optical depth is proportional to the
column density, the templates can be adjusted for any column density
by multiplying by $N/10^{21}$. The absorption spectrum is then computed
as $e^{-\tau_i}$ where $\tau_i$ is the value of the scaled optical depth 
template at each wavelength. Rotational transitions $J^{\prime\prime}>8$ are not
seen in the data and thus we exclude them from the modeling.

We perform the model fits using our custom MCMC routine based on the algorithm
of \citet{goodman10} \citep[see also][]{foreman13}. Uniform 
priors are assumed for all parameters. We run the MCMC for 10,000 steps with
100 walkers, resulting in 10$^6$ samplings of the posterior distribution.
Because the template spectra are calculated for discrete values of $b$ we
interpolate between the two templates which bracket the model value of $b$. For
example, if the MCMC iteration requires a spectrum with $b = 3.2$ km s$^{-1}$
we linearly interpolate between the $b = 3.0$ km s$^{-1}$ and $b = 4.0$ km s$^{-1}$
templates. Once the interpolation is complete the resulting spectrum is shifted
by the radial velocity and scaled to find the best-fit column density.  The model
spectrum, which is at the native velocity resolution, is convolved with the COS
LSF for comparison with the observed spectrum. The median values of the
marginalized posterior distributions are taken as the best-fit parameter values
and the $1\sigma$ confidence intervals are the 16$^\text{th}$ and 84$^\text{th}$
percentiles of the marginalized posteriors. 

\begin{figure*}[htbp]
   \centering
   \includegraphics[scale=.80,clip,trim=5mm 15mm 35mm 20mm,angle=0]{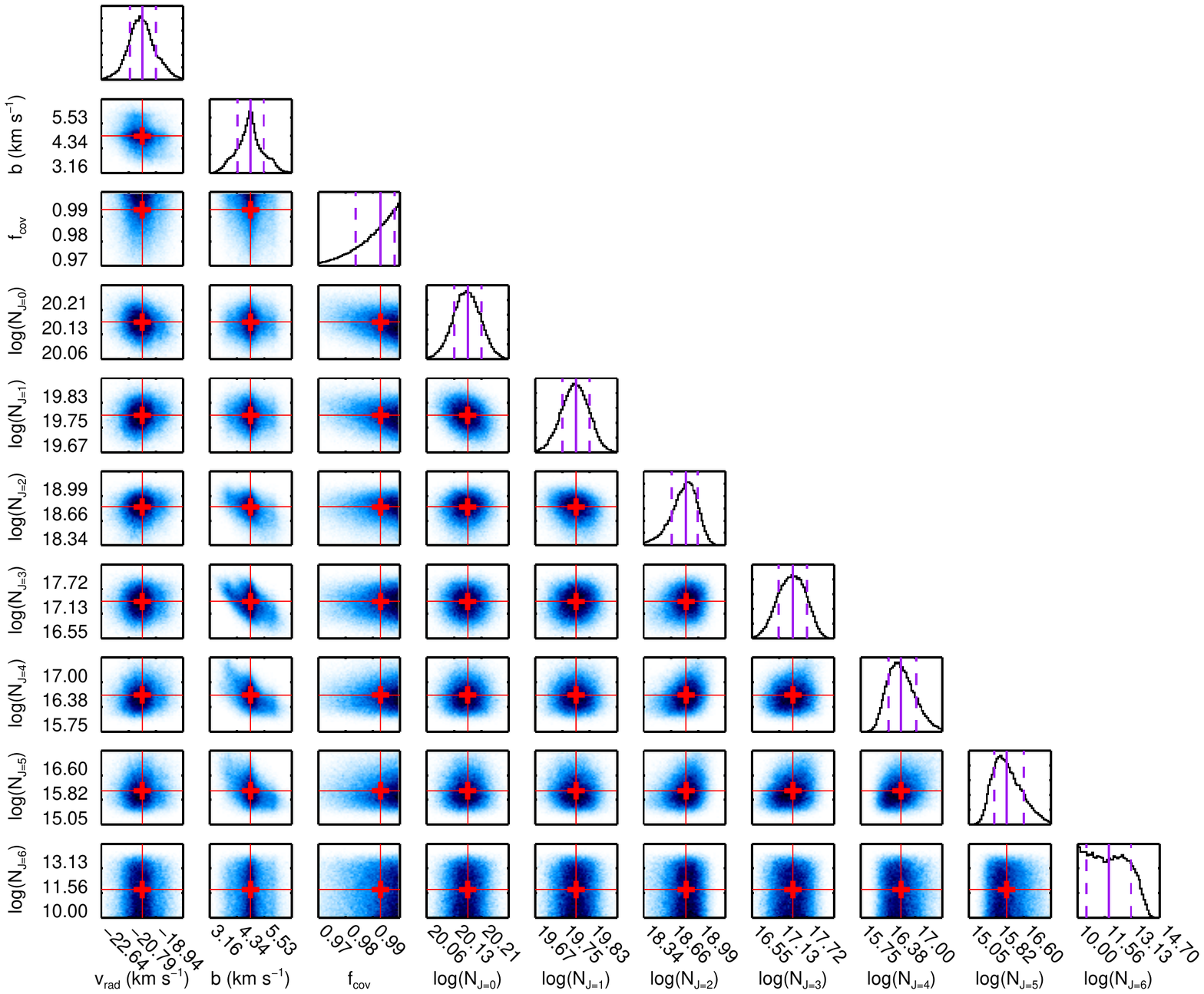}

   \figcaption{Corner plot of the marginalized posterior distributions for the
H$_2$ MCMC of HK Ori.  Note that all column densities $N$ are in units of
cm$^{-2}$.  \label{fig:h2corner_hk}}

\end{figure*}

\begin{figure*}[htbp]
   \centering
   \includegraphics[scale=.80,clip,trim=5mm 15mm 35mm 20mm,angle=0]{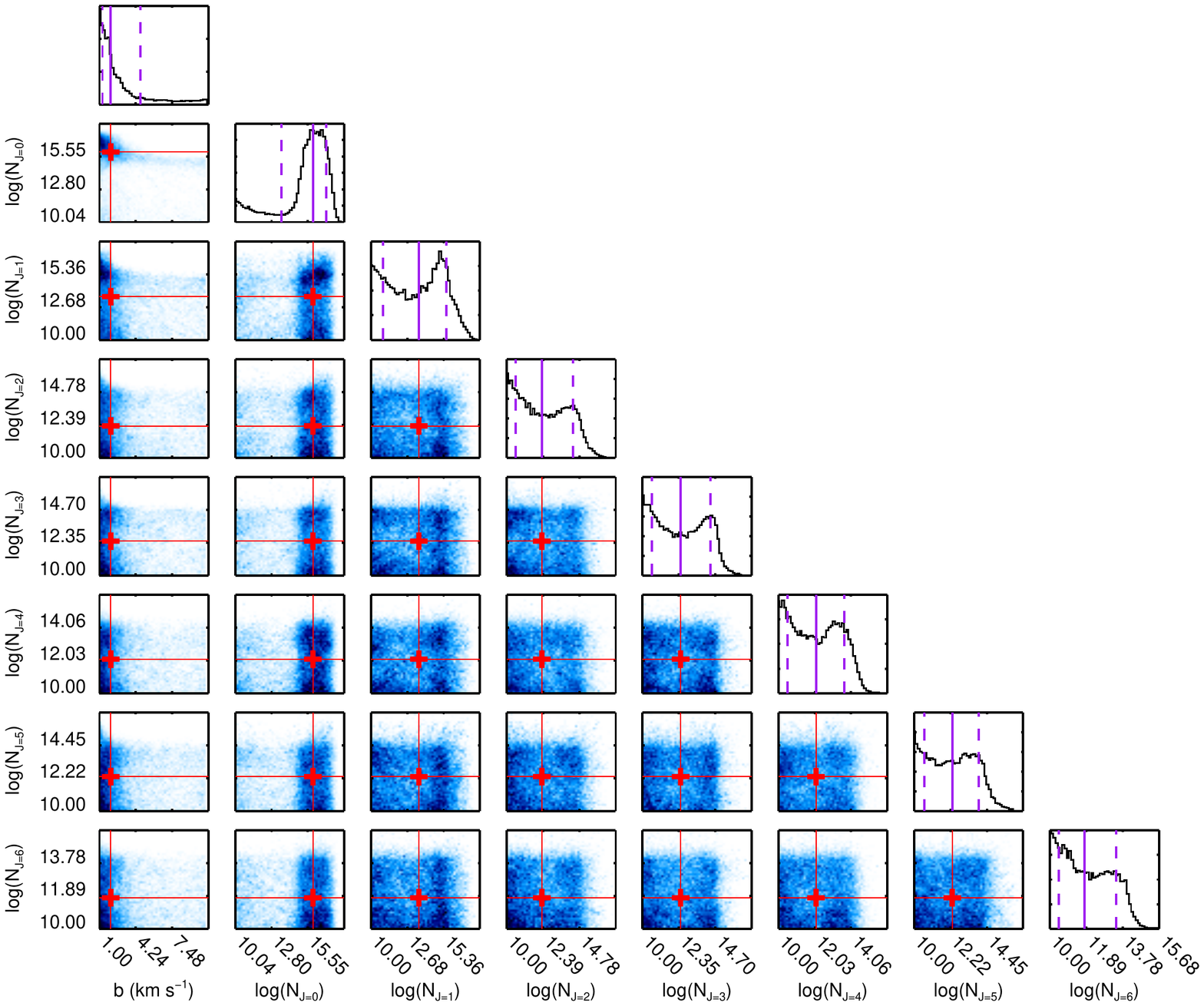}

   \figcaption{Corner plot of the marginalized posterior distributions for the
H$_2$ MCMC of T Ori. The parameters are almost entirely unconstrained due to
the lack of absorption in the observed spectrum. \label{fig:h2corner_t}}

\end{figure*}

\begin{figure*}[htbp]
   \centering
   \includegraphics[scale=.70,clip,trim=5mm 15mm 5mm 30mm,angle=0]{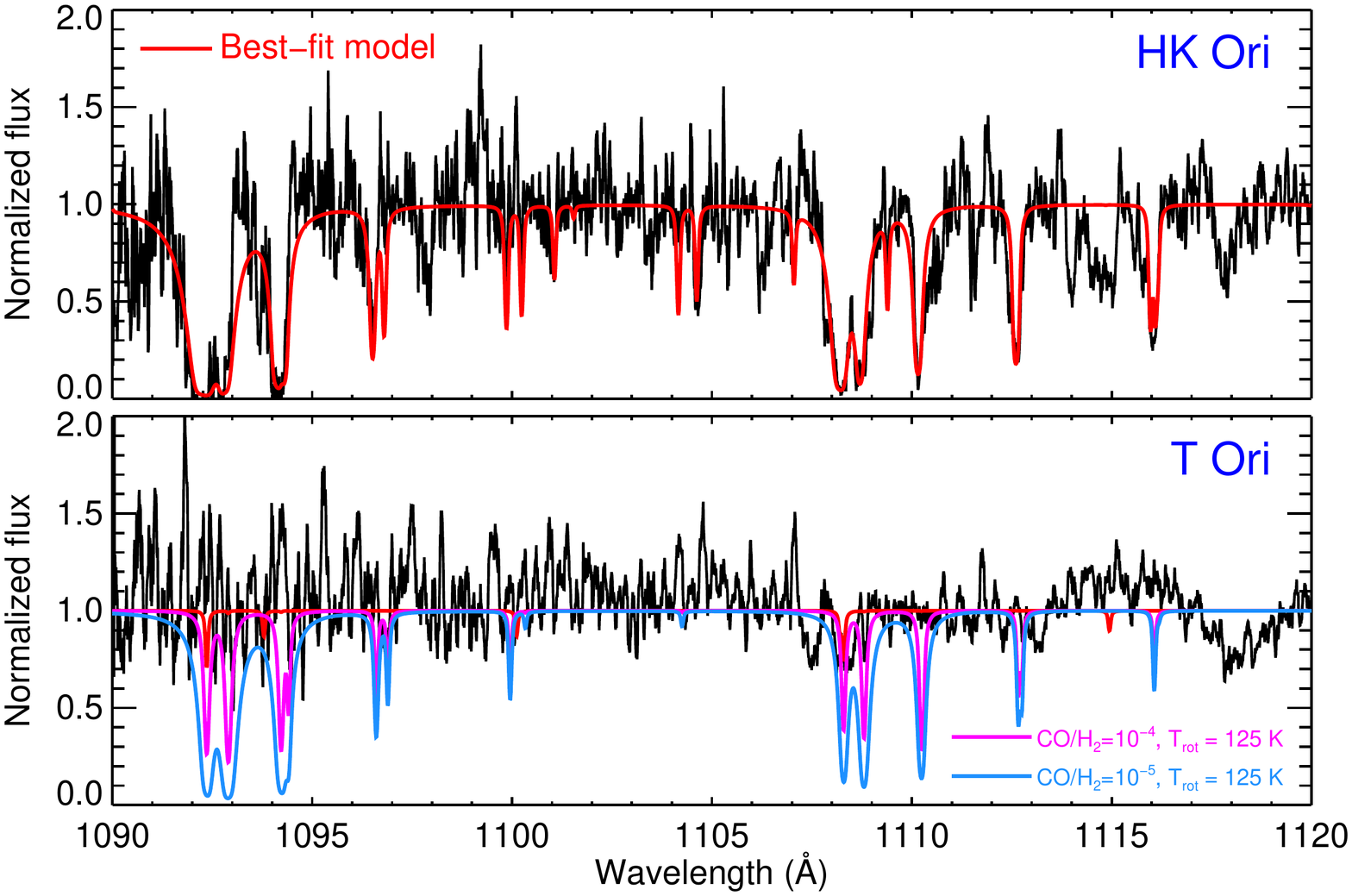}

   \figcaption{Normalized \hst\ COS spectra of HK Ori (top) and T Ori (bottom) from 1090
   -- 1120 \AA. The best-fit \htwo\ absorption models are over-plotted in red. No
significant \htwo\ absorption is detected in the spectrum of T Ori. The magenta model
shows the expected \htwo\ absorption assuming a conservative CO/\htwo\ ratio of
10$^{-4}$ and the same temperature as that derived for the CO gas (see \autoref{sec:comodels}).
The blue model shows an assumed ratio of 10$^{-5}$.
\label{fig:h2mods}}

\end{figure*}

For T Ori there is no visual evidence of H$_2$ absorption (see
\autoref{fig:h2specs} and \autoref{fig:h2mods}) and the spectrum 
from $1090 - 1130$ \AA\ is almost pure
continuum (surprising given the strong CO detection, discussed in the following section). 
We apply the above H$_2$ fitting procedure to T Ori's spectrum, with some exceptions, to 
measure the upper limit to T Ori's  H$_{2}$ absorption properties. For T Ori we fix the
coverage fraction at $f_\text{cov} = 1.0$ and we freeze the radial velocity
near a literature value for the system of $v_\text{rad} = 56.0$ km s$^{-1}$
\citep{cauley15a}.  This is necessary because the lack of absorption lines in
the spectrum allows a total degeneracy between $f_\text{cov}$ and the H$_2$
column densities. For the same reason $v_\text{rad}$ is entirely unconstrained.

The marginalized posterior distributions from the MCMC are shown for HK Ori in
\autoref{fig:h2corner_hk} and for T Ori in \autoref{fig:h2corner_t}. Darker
blue colors indicate regions of higher posterior density. The fixed parameters
for T Ori $f_\text{cov}$ and $v_\text{rad}$ are excluded from
\autoref{fig:h2corner_t}. We also exclude the posteriors for the rotational
states $J^{\prime\prime} = 7,8$ for both objects due to severe blending of the
$J^{\prime\prime} = 7$ state and no detectable features corresponding to the
$J^{\prime\prime} = 8$ state. 

The most likely model parameters and their $1\sigma$ confidence intervals are
given in \autoref{tab:h2fits} and the best-fit models are shown in
\autoref{fig:h2mods}. As expected, we are only able to place an upper
limit on the \htwo\ column density for T Ori due to the lack of any discernible H$_2$
absorption in the spectrum. Also over-plotted for T Ori in \autoref{fig:h2mods}
is the expected \htwo\ absorption assuming the canonical interstellar dense and translucent cloud ratios (CO/\htwo$=10^{-4}$ and $=10^{-5}$)
and temperature $T_\text{rot} = 125$ K where the temperature and CO column
density are those derived in \autoref{sec:comodels}. A model with such a large
\htwo\ column density is strongly ruled out by the MCMC posteriors. We will return to potential
explanations for the lack of circumstellar H$_2$ absorption around T Ori in \autoref{sec:disc}. 

\begin{deluxetable}{lcc}
\tablewidth{0pt}
\tablecaption{H$_2$ model fit parameters\label{tab:h2fits}}
\tablehead{\colhead{Parameter}&\colhead{HK Ori}&\colhead{T Ori}}
\colnumbers
\tabletypesize{\scriptsize}
\startdata
$N$($J''=0$)$^\dagger$ (cm$^{-2}$) & 20.17$^{+0.04}_{-0.04}$ & 15.8$^{+1.0}_{-2.8}$ \\
$N$($J''=1$) (cm$^{-2}$) & 19.80$^{+0.04}_{-0.04}$ & $<$15.5 \\
$N$($J''=2$) (cm$^{-2}$) & 18.85$^{+0.13}_{-0.17}$ & $<$14.3 \\
$N$($J''=3$) (cm$^{-2}$) & 17.44$^{+0.30}_{-0.31}$ & $<$14.4 \\
$N$($J''=4$) (cm$^{-2}$) & 16.75$^{+0.35}_{-0.30}$ & $<$13.7 \\
$N$($J''=5$) (cm$^{-2}$) & 16.12$^{+0.48}_{-0.35}$ & $<$13.9 \\
$N$($J''=6$) (cm$^{-2}$) & 11.71$^{+1.33}_{-1.23}$ & $<$13.5 \\
$N$($J''=7$) (cm$^{-2}$) & \nodata & \nodata \\
$N$($J''=8$) (cm$^{-2}$) & 12.58$^{+1.59}_{-1.85}$ & $<$15.5 \\
$v_\text{rad}$ (km s$^{-1}$) & -19.9$^{+0.9}_{-0.9}$ & 56.0 \\
$b$ (km s$^{-1}$) & 4.9$^{+0.58}_{-0.60}$ & 2.1$^{+3.8}_{-0.8}$ \\
$f_\text{cov}$ & 0.995$^{+0.004}_{-0.007}$ & 1.0 \\
$T_\text{rot}$ (K) & 141$^{+6}_{-6}$ & 452$^{+649}_{-178}$ \\
\enddata
\tablenotetext{\dagger}{All column densities $N$ are log$_{10}$ values.}
\end{deluxetable}

The most likely parameter values for HK Ori, on the other hand, give a total
H$_2$ column density of log$_{10}$(N[H$_2$])$=20.34^{+0.03}_{-0.03}$ cm$^{-2}$.
The broadening parameter value is $b = 4.9^{+0.58}_{-0.60}$ km s$^{-1}$ suggesting a
significant contribution from turbulence or unresolved components to the line widths. 
The model also indicates that the H$_2$ line-of-sight absorption covers the entire stellar
disk given the value of $f_\text{cov}$ so close to 1.0.

The temperature of the H$_2$ is not a free parameter in the spectrum fitting.
Instead, we use the derived column densities to find the most-likely rotational
temperature. The low-$J^{\prime\prime}$ H$_2$ level populations in the disk are expected to be determined by
collisions so we can calculate the temperature of the gas assuming a
Maxwell-Boltzmann distribution of the rotational states
\citep[see][]{france14}. We accomplish this by using the same MCMC routine to
fit a line to the H$_2$ densities as a function of the excitation temperature
of the various rotational states. The rotational, or kinetic, temperature
$T_\text{rot}$ of the gas is then the slope of the best-fit line. 

\begin{figure}[htbp]
   \centering
   \includegraphics[scale=.4,clip,trim=10mm 10mm 5mm 25mm,angle=0]{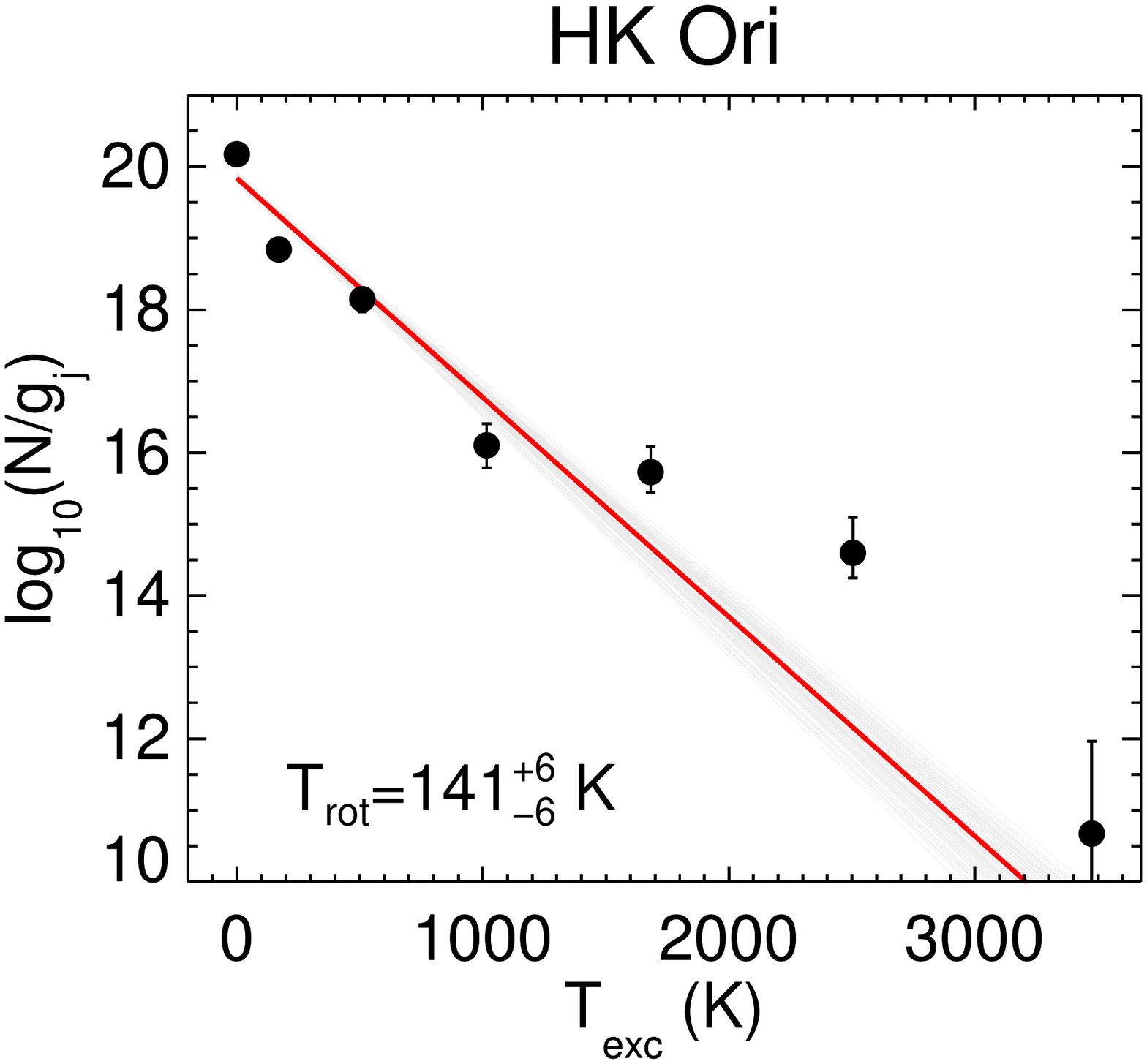}

   \figcaption{Fit to the \htwo\ excitation temperature versus column density
   for HK Ori. One hundred random samplings of the posterior distributions are shown in gray.
The best-fit value for the rotational excitation temperature is $T_\text{rot} =
141^{+6}_{-6}$ K, indicative of material in a warm disk layer.
\label{fig:hkori_tempfit}}

\end{figure}




The most-likely linear fit is shown with a red line in \autoref{fig:hkori_tempfit}
along with one hundred models from a random sampling of 
the accepted posterior distributions (gray lines). We find that the H$_2$ gas 
around HK Ori has a rotational temperature of $T_\text{rot} = 142^{+6}_{-6}$ K. 
The small fit uncertainties on the $J^{\prime\prime} = 0,1,2$ dominate the slope fit 
and thus the temperature determination. As expected from the lack of any \htwo\ 
absorption around T Ori, the corresponding temperature fit is mostly unconstrained:
the same procedure yields $T_\text{rot} = 452^{+649}_{-178}$ K for T Ori. 

\subsection{CO models}
\label{sec:comodels}

The CO fitting procedure is outlined in detail in \citet{mcjunkin13} \citep[see
also][]{france14} and we briefly summarize the salient points here. To derive
the circumstellar CO column densities we fit the CO Fourth Positive band system
shaded in gray in \autoref{fig:cospecs}. We include the $(4-0), (3-0),
\text{and} (2-0)$ bands in the fits for both objects but exclude the $(1-0)$
band for the HK Ori fit due to contamination by what is most likely an emission
line from atomic iron. The synthetic CO spectra are generated using literature
oscillator strengths and ground-state energy levels
\citep{haridass94,eidelsberg99,eidelsberg03}. The free parameters in the MCMC
fitting procedure are the column densities of $^{12}$CO and $^{13}$CO, the
radial velocity of the absorbing gas $v_\text{rad}$, the rotational temperature
of the gas $T_\text{rot}$, and the Doppler-broadening parameter $b$. Similar
to the case for H$_2$, $b$ has a strong affect on the optical depth of the CO
transitions and is somewhat degenerate with the column density. This makes
the value of $b$ sensitive to not only the line widths but the line depths as well.

We again adopt uniform, non-restrictive priors for most of the parameters and run the 100
walkers for 10,000 steps each. The one exception is the Doppler broadening
parameter $b$: we restrict the prior range to $0.0 < b \leq 5.0$ km s$^{-1}$.
For each model iteration we generate the CO spectrum using the input $T_\text{rot}$, 
$b$, and column density values. The spectrum is then shifted in velocity space 
according to $v_\text{rad}$ and then convolved with the COS LSF to produce the output model. 

\begin{figure}[htbp]
   \centering
   \includegraphics[scale=.40,clip,trim=15mm 15mm 35mm 20mm,angle=0]{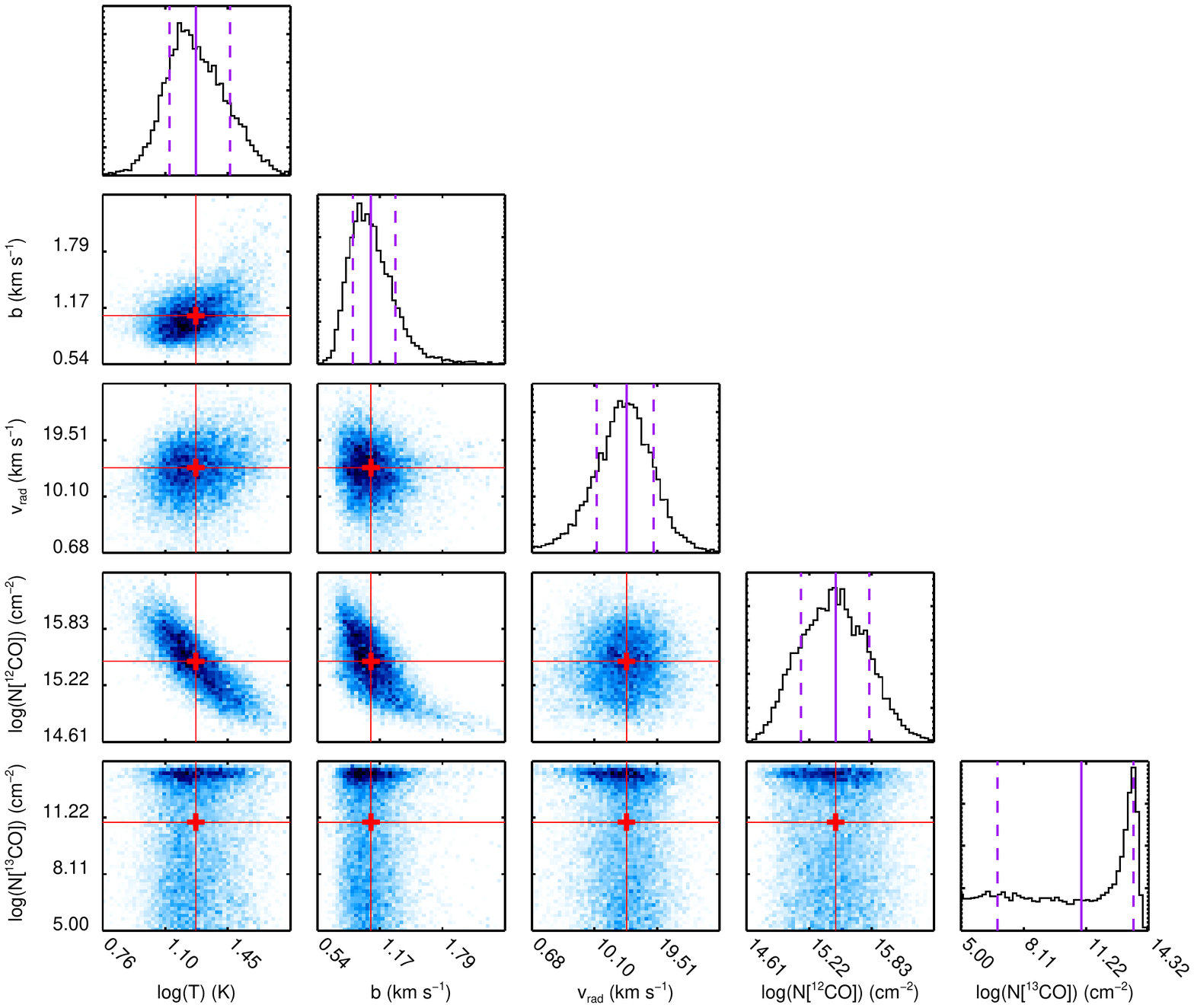}

   \figcaption{Corner plot of the marginalized posterior distributions for the CO MCMC of HK Ori.
\label{fig:cocorner_hk}}

\end{figure}

\begin{figure}[htbp]
   \centering
   \includegraphics[scale=.40,clip,trim=15mm 15mm 35mm 10mm,angle=0]{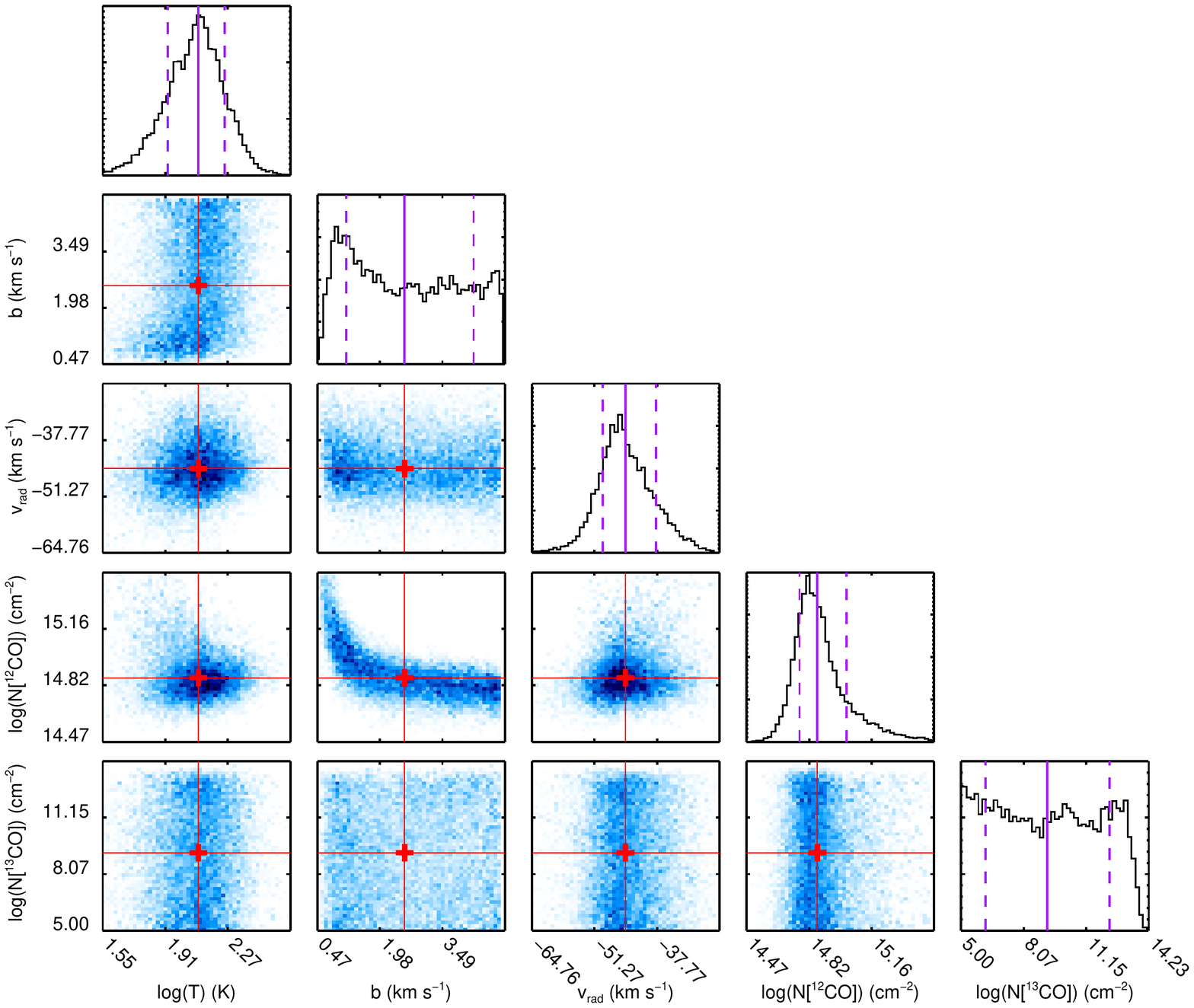}

   \figcaption{Corner plot of the marginalized posterior distributions for the CO MCMC of T Ori.
\label{fig:cocorner_t}}

\end{figure}

\begin{figure*}[ht!]
   \centering
   \includegraphics[scale=.60,clip,trim=5mm 5mm 10mm 60mm,angle=0]{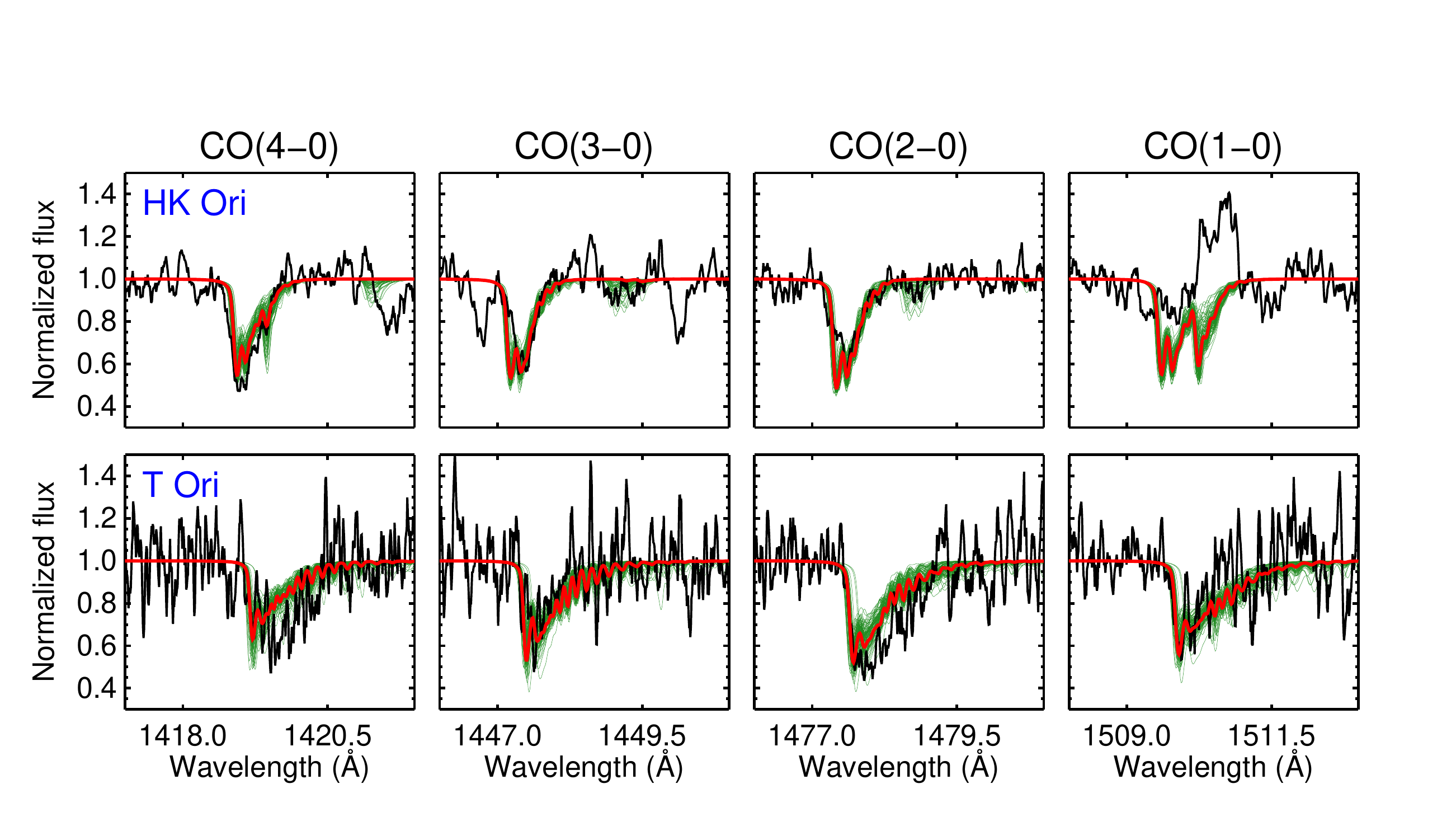}

   \figcaption{Normalized spectra surrounding the CO (\textit{A-X}) absorption
bands for HK Ori (top row) and T Ori (bottom row). The best-fit CO absorption
models are over-plotted in red and one hundred random draws from the accepted
posterior chains are shown in green. Note that the CO(1-0) band is excluded
from the model determination for HK Ori due to the contamination by the strong
emission line near 1511 \AA. \label{fig:comods}}

\end{figure*}

The marginalized posterior distributions for the CO MCMC fits are shown in
\autoref{fig:cocorner_hk} and \autoref{fig:cocorner_t}. The best-fit models
(red) and fifty random samplings of the accepted posteriors (green) are shown
in \autoref{fig:comods}. The most noticeable difference in the CO spectra is
the width of the absorption bands, where HK Ori's are fairly narrow while T
Ori's are broad. This is manifested in the large difference in derived
$T_\text{rot}$ values: for HK Ori $T_\text{rot} = 19^{+11}_{-6}$ K and for T
Ori $T_\text{rot} = 124^{+53}_{-44}$ K. Note that T Ori's $b$-value
is just the median of the prior range, i.e., it is unconstrained. This is
a result of the high temperature: as the temperature increases and the optical
depth in the higher rotational states increases, the individual line broadening
becomes less important and the instrumental profile makes differences in
$b$ negligible. This is not the case for the low temperature fits for HK Ori.
At low temperatures larger values of $b$ increase the optical depth over
a narrow range of rotational states which significantly increases the absorption
depth. Thus there is some degeneracy between $N$ and $b$ at low $T$ and
$b$ is constrained even though it is much smaller than the instrumental resolution.
Neither spectrum shows any strong absorption due to $^{13}$CO, in contrast to some
CTTS disks \citep[e.g.][]{france12}.

We note that the $^{12}$CO/$^{13}$CO ratios indicated by our analysis
are much larger than is typically found for the ISM or the circumstellar 
environments of young stellar objects (YSOs) \citep{smith15}. However, the
$^{13}$CO column densities are highly uncertain and it's likely that the
$^{13}$CO absorption is almost entirely lost in the noise for both objects.
Thus we caution against applying the derived $^{12}$CO/$^{13}$CO ratios
to further work on these systems.

\begin{deluxetable}{lcc}
\tablecaption{CO model fit parameters\label{tab:cofits}}
\tablehead{\colhead{Parameter}&\colhead{HK Ori}&\colhead{T Ori}}
\colnumbers
\tabletypesize{\scriptsize}
\startdata
$N(^{12}\text{CO})^\dagger$ (cm$^{-2}$) & 15.5$^{+0.3}_{-0.3}$ & 14.9$^{+0.2}_{-0.1}$ \\
$N(^{13}\text{CO})$ (cm$^{-2}$) & 11.0$^{+2.6}_{-4.2}$ & 9.2$^{+3.1}_{-3.0}$ \\
$v_\text{rad}$ (km s$^{-1}$) & $-$14.9$^{+4.1}_{-4.6}$ & 44.5$^{+6.7}_{-5.1}$ \\
$b$ (km s$^{-1}$) & 1.1$^{+0.3}_{-0.2}$ & 2.5$^{+1.7}_{-1.4}$ \\
$T_\text{rot}$ (K) & 19$^{+11}_{-6}$ & 124$^{+53}_{-44}$ \\
\enddata
\tablenotetext{\dagger}{Column densities $N$ are log$_{10}$ values.}
\end{deluxetable}

\subsection{\ion{H}{1} Absorption}
\label{sec:HImodels}

\begin{figure}[htbp]
   \centering
  \includegraphics[scale=.38,clip,trim=25mm 10mm 10mm 10mm,angle=0]{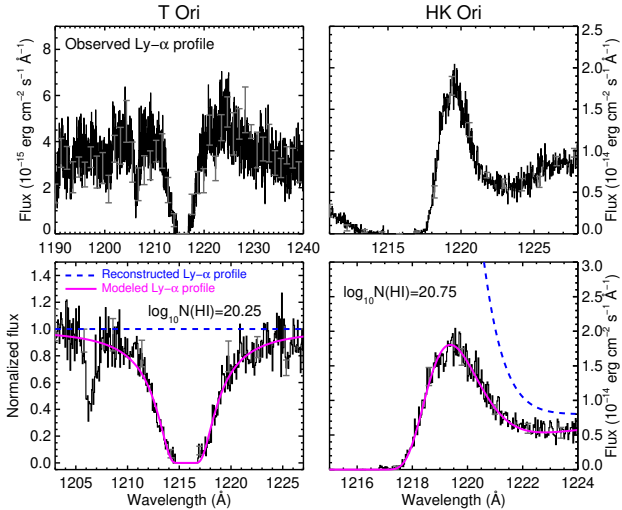}

   \figcaption{Observed (top panels) and reconstructed (bottom panels) Ly$\alpha$ profiles for T Ori and HK Ori. The best-fit neutral hydrogen ISM column densities are labeled in the bottom panels. \label{fig:lymmods}}

\end{figure}

Absorption line analyses of interstellar sight-lines often rely on a knowledge of the 
total neutral hydrogen column density, N(\ion{H}{1}), to place molecular absorption 
properties in context.  Toward this end, we use the 2012 $HST$-COS G130M observations 
of both stars to measure N(\ion{H}{1}) by fitting the Ly$\alpha$ absorption profile. 
We show the observed profiles and reconstructed spectra from the model fits in \autoref{fig:lymmods}. 
Both HK  Ori and T Ori display broad Ly$\alpha$ emission features consistent with 
accretion-generated \ion{H}{1} emission observed on lower mass T Tauri stars
\citep[e.g.,][]{france14b}. In analogy to the methodology employed for H$_{2}$ and CO, 
we assume the observed profile is the superposition of an underlying stellar emission
spectrum and the Ly$\alpha$ absorber.  For HK Ori, we employed a broad Gaussian as 
the intrinsic Ly$\alpha$ profile shape, and implemented an iterative fitting routine 
to find the best fit to the observed COS G130M Ly$\alpha$ spectra \citep{france12,mcjunkin13}.  
The Ly$\alpha$ emission peak is less pronounced in T Ori, and here we fit a spline function 
to the continuum and emission line as a baseline for fitting the \ion{H}{1} absorption. 
The Doppler $b$-value, $b_{HI}$, is assumed to be 10 km s$^{-1}$ for both stars. For 
more details on reconstructing intrinsic Ly$\alpha$ profiles see \citet{mcjunkin14}. 

For HK Ori, our fits find a baseline continuum of 8~$\times$~10$^{-15}$ erg 
cm$^{-2}$ s$^{-1}$ \AA$^{-1}$, Gaussian peak amplitude of 6~$\times$~10$^{-12}$ 
erg cm$^{-2}$ s$^{-1}$ \AA$^{-1}$, a Gaussian FWHM = 1100 km s$^{-1}$, and 
log$_{10}$N(\ion{H}{1}) = 20.75~$\pm$~0.10. Taking the  spline function as the 
underlying stellar spectrum for T Ori, we find log$_{10}$N(\ion{H}{1}) = 
20.25~$\pm$~0.10. These results are summarized in Table 3.  

\section{DISCUSSION}
\label{sec:disc}

\subsection{Comparing H$_{2}$, CO, and \ion{H}{1} with the Translucent ISM}
\label{sec:ismcomp}

We compared the absorption properties of the two targets against typical properties 
of interstellar sight-lines to help solidify assignment of these components to 
a circumstellar disk origin. The HK Ori CO/H$_{2}$ ratio and molecular 
fraction\footnote{The molecular fraction, $f_{H2}$ is defined as 
$f_{H2}$~=~2N (H$_{2}$)/(2N(H$_{2}$) + N(\ion{H}{1})).} are consistent with interstellar clouds 
at the diffuse-to-translucent boundary \citep[e.g.,][]{burgh07}. However, the CO 
and molecular fraction are considerably higher than expected for a star with HK 
Ori's optical reddening $E(B - V)$ \citep[$=0.12$; assuming the N(\ion{H}{1}) to 
reddening conversion from][]{diplas94}. Furthermore, the H$_{2}$ column density 
to HK  Ori is orders of magnitude higher than what is seen to typical hot star 
targets near the Orion star-forming region \citep{savage77}. This points to a 
circumstellar origin of the molecular gas in HK Ori.

Further support for a circumstellar origin of the molecular gas around HK  Ori comes from an 
analysis  of the excitation temperatures.  While T(H$_{2}$) = 141 K is consistent with an
interstellar origin, T(CO) = 19 K is much higher than what is seen for translucent clouds
\citep[e.g., see Figure 6 of][]{burgh07}. We note that the temperature discrepancy could be
due to the density of CO being below the critical density (see \autoref{sec:ismcomp} for
a more detailed explanation). Therefore, numerous lines of evidence suggest a
circumstellar, as opposed  to  interstellar, origin for the molecular gas observed toward HK  Ori.  

T Ori's hydrogen sightline is characteristic of hot stars near the Orion star 
forming regions \citep[N(H$_{2}$)~$<$~10$^{16}$ cm$^{-2}$; N(\ion{H}{1}) = 10$^{20.0-20.5}$ 
cm$^{-2}$;][]{savage77}. However, the diffuse sightlines in Orion are typically 
associated with very low CO upper limits \citep[N(CO)~$<$~10$^{13}$ cm$^{-2}$;][]{federman80}.
Similarly, the very low molecular fraction is incompatible with the large CO column density, 
arguing that the CO we observe towards T Ori is almost certainly of a circumstellar origin.  
This is corroborated by the high T(CO), which is a factor of $\sim$10~--~20 higher than 
seen on translucent interstellar sight-lines \citep{burgh07,sheffer08}. The circumstellar
nature of T Ori's CO absorption is also bolstered by the fact that it shows a variable
column density likely related to inner disk material (see \autoref{sec:tori_h2}).

\begin{deluxetable*}{lccccc}
\tablecaption{Total column densities and N(CO)/N(H$_2$) ratios\label{tab:ratio}}
\tablehead{\colhead{Object}&\colhead{N(H$_2$)}&\colhead{N(CO)}& \colhead{N(\ion{H}{1})}& \colhead{N(CO)/N(H$_2$)} & \colhead{$f_{H2}$}}
\colnumbers
\startdata
HK Ori & 20.34$^{+0.03}_{-0.03}$ & 15.5$^{+0.5}_{-0.2}$ & 20.75~$\pm$~0.10  & 1.3$^{+1.6}_{-0.7} \times 10^{-5}$ &  0.44  \\
T Ori & $<$~15.9$^{+1.0}_{-0.4}$ & 14.9$^{+0.2}_{-0.1}$ & 20.25~$\pm$~0.10 &    \nodata  &     $<$1~$\times$~10$^{-4}$  \\
\enddata
\end{deluxetable*}


\subsection{Where's the H$_2$ around T Ori?}
\label{sec:tori_h2}

The lack of H$_{2}$ absorption towards T Ori is surprising given the system's
young age and the high likelihood that we observe the disk closer to an edge-on
orientation. The properties of the CO gas we derive confirm that the gas is 
circumstellar ($T \approx 125$ K) and not interstellar, where the gas typically 
has temperatures $T < 10$ K. Furthermore, the radial velocity of the CO absorption, 
$\approx 45$ km s$^{-1}$, is consistent with, although mildly blue-shifted relative 
to, the system velocity of $\approx 56$ km s$^{-1}$ (noting that $HST$-COS has a 
wavelength accuracy of 15 km s$^{-1}$). 


\begin{figure*}[htbp]
   \centering
  \includegraphics[scale=.75,clip,trim=15mm 10mm 10mm 20mm,angle=0]{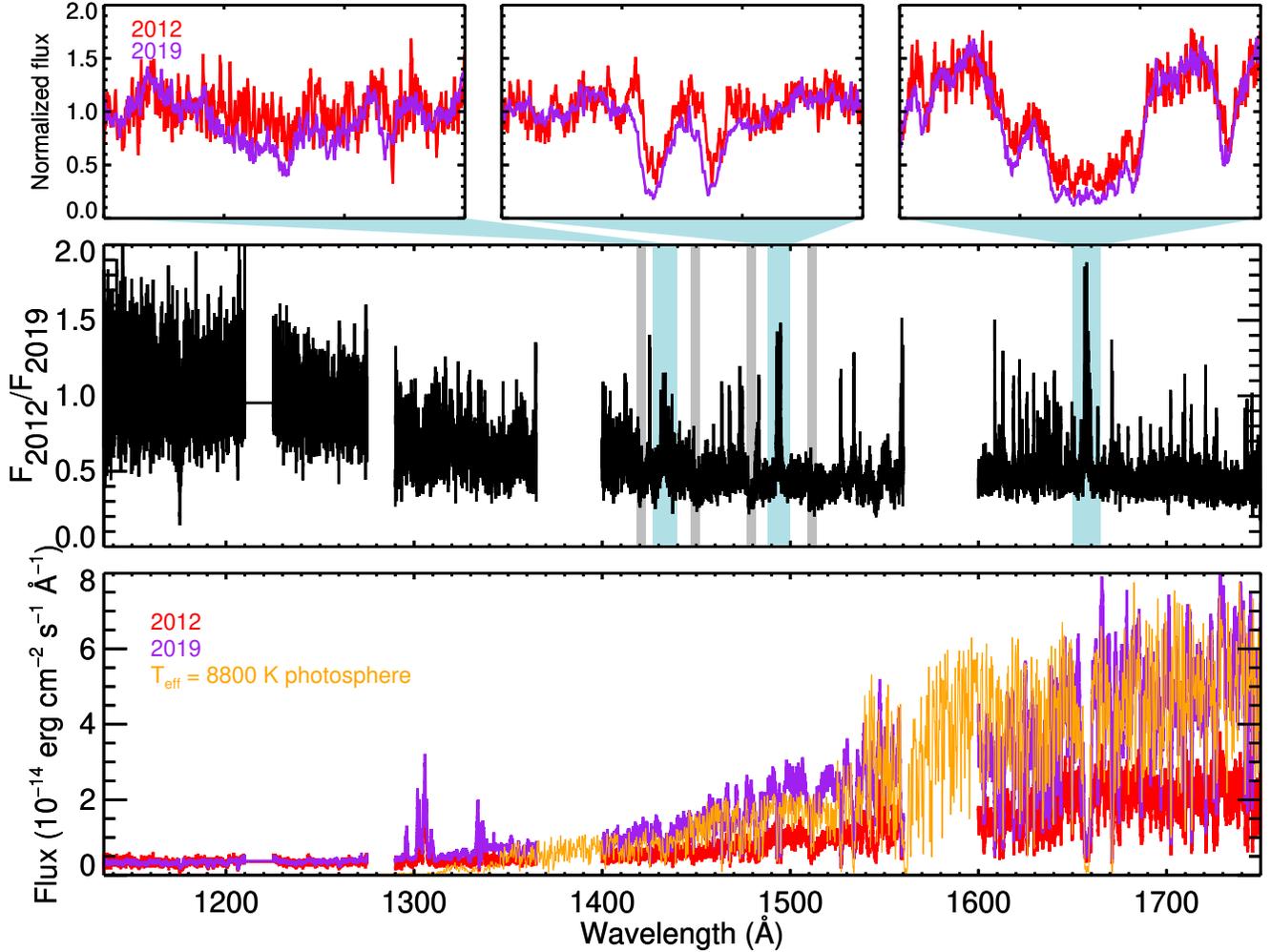}

   \figcaption{Comparison of the 2012 (red) and 2019 (purple) FUV T Ori spectra. 
   The bottom panel displays the flux spectra from both epochs, as well as a 
   photospheric template scaled to match the 2019 flux. The Lyman-$\alpha$ emission
   lines are interpolated over for clarity. Only wavelengths common to both
   sets of observations are shown. The middle panel shows the ratio of the 2012
   spectrum to the 2019 spectrum and the top panels show zoomed-in normalized
   spectrum comparisons between strong spectral features from each epoch. The CO
   bandheads are marked with gray shaded regions and the wavelength ranges shown
   in the top panels are marked with light blue shading in the middle panel.
   \label{fig:toriflux}}

\end{figure*}

We can look to the variability between the 2012 and 2019 observations for
clues about the lack of H$_2$ absorption towards the star. In \autoref{fig:toriflux}
we show a comparison between the \textit{HST} COS spectra from 2012 (red)
and 2019 (purple). The bottom panel shows the raw spectra for wavelengths
common to both epochs and the middle panel shows the ratio of the fluxes. The
top panel displays zoomed-in normalized comparisons between strong spectral
features. A few things are evident from \autoref{fig:toriflux}: 1. The flux
at wavelengths longer than $\approx 1300$ \AA\ was $\approx 2\times$ greater
in 2019 than in 2012; 2. The flux at wavelengths shorter than $\approx 1300$ \AA\
stayed relatively constant; 3. Most of the photospheric absorption lines are weaker
when the star is fainter; 4. The CO absorption is stronger when the star is
fainter (see the flux dips within the gray bands in the middle panel of
\autoref{fig:toriflux}). Additionally, we downloaded $V$-mag ASAS-SN light curves
\citep{shappee14,kochanek17} of T Ori and interpolated the observed magnitudes
onto our \textit{HST} COS observation dates: for the 2012 observations $V \approx 11.3$
and for the 2019 observations $V \approx 10.4$, which translates to a factor
of $\approx 2.3$ in flux. The $V$-band measurements confirm that the optical
flux varies at a similar level to the FUV flux.

We speculate that T Ori's nature as a rapid rotator and UXOR variable 
can qualitatively explain the phenomena seen in \autoref{fig:toriflux}. 
First, rapidly rotating stars are subject to non-negligible gravity darkening 
where the oblateness of the star causes the poles to be hotter than the 
equator \citep{vonzeipel24,espinosa11}. If the UXOR screen preferentially blocks 
equatorial stellar latitudes, or is at least more optically thick across these 
regions, the star will be dimmer but
the stellar spectrum will be weighted towards the hotter polar regions. This
can account for the observed shallower absorption lines in the 2012 faint phase.
Second, the increased CO absorption during the faint phase could be a result of the
increased gas column density produced by the occulting screen. Finally, the lack
of H$_2$ absorption in the Lyman transitions may be the result of the
sub-1300 \AA\ flux, which is related to material accreting onto the star, being 
scattered into the line-of-sight over a large volume in the inner disk region. 
If the flux is the result of scattering over a large volume, we are not probing 
the same pencil beam through the disk that produces the CO absorption since the 
flux from 1400 - 1500 \AA\ is largely photospheric in origin. This also explains 
the relatively stable sub-1300 \AA\ flux since the UXOR screen does not occult 
the entire scattering surface.

We stress that the scenario outlined above is speculative; there are
likely other plausible explanations for the various phenomena (e.g., an occulting
accretion column and related veiling of the stellar photosphere). In particular, the 
proposed viewing geometry must be fine-tuned in order to preferentially occult 
equatorial latitudes on the star, leaving the hotter poles visible. Additional 
FUV time-series spectra will be useful in determining a more precise view of the 
system geometry and we defer a more quantitative characterization of the variability 
to future work.

\subsection{Large Doppler broadening parameters}
\label{sec:doppler}

The H$_2$ and CO analysis for HK Ori suggest large
($\approx 1-4$ km s$^{-1}$) non-thermal broadening components are necessary to
reproduce the observed H$_2$ and CO absorption. This would imply significant contributions
from turbulence or spatially unresolved gas motions responsible for the line broadening.
The turbulence interpretation seems unlikely given that CO radio observations suggest 
that turbulence in protoplanetary disks is weak at distances of $\approx 10s$ of AU 
\citep{hughes11,flaherty18,flaherty20}. On the other hand, the pencil beam H$_2$ observations, 
if passing through the disk, likely probe higher surface layers than the thermal CO emission.

We caution against an over-interpretation of the large value for 
the Doppler parameter for two reasons. First, absorption features with widths of a
few kilometers per second are not resolved in our spectra. The large $b$-values are
mainly a result of the models fitting the absorption line depths, which are
determined by both $b$ and the column density $N$. Without resolving the line widths
of individual H$_2$ and CO transitions it is difficult to say with confidence
what is driving the large values of $b$.
Second, there is considerable uncertainty around the viewing geometry for the
disk around HK Ori. More precise constraints on
our viewing angle through these circumstellar environments
could help clarify the interpretation of the large value for the Doppler parameter. 

\subsection{Spatial Location and Distribution of Molecular Absorption Components around HK Ori}
\label{sec:ismcomp}

We have measured significant columns of H$_2$ and CO towards HK Ori. The
presence of the K4 companion at $0\overset{\arcsec}{.}3$ might suggest that the COS
observations contain flux from both components. While it is plausible that the
less massive companion contributes some coronal emission line flux to the FUV spectrum,
the absorption signatures are most likely produced by circumstellar material
around the primary: \citet{smith05} found no evidence for an IR excess around
the K4 companion whereas the IR excess around HK Ori A is well matched by
an accretion disk extending to $\approx 30$ AU. Given the lack of circumstellar
material around the secondary, we continue under the assumption that the absorption
is created by material occulting the primary star.

The most critical assumption necessary for deriving a CO/\htwo\ ratio in
a disk via pencil beam absorption is that the CO and \htwo\ populations are
co-spatial, i.e., we are sampling the same parcels of the disk in both
species. At first glance, the differences in the derived rotational temperatures between
the H$_{2}$ and CO populations suggest we might be sampling different gas components.  
However, while the CO rotational levels are populated by collisions, radiative processes 
can alter the observed distributions of $J$-levels. If the local number density is lower 
than the CO critical density, these levels can radiatively depopulate fast enough that the 
populations are not representative of the local kinetic temperature. 

Referencing the density and temperature structures of Herbig Ae disks presented in \citet{agundez18}, 
we estimate the local H number density in the region probed by our HK Ori H$_{2}$ absorption spectra is
$n_{H}$~$\sim$~10$^{4-7}$ cm$^{-3}$. Assuming an H$_{2}$ ortho–para ratio
of 3 and that the T$_{H2}$~$\approx$~T$_{kin}$ = 150 K, the total ortho- + para-H2 
collision rate, summed over all possible lower levels, is $\Gamma_{TOT}$ = 2.6 $\times$ 
10$^{-10}$ cm$^{3}$ s$^{-1}$ at $J = 10$ \citep{yang10}. The CO critical density for $J = 10$ at 150 K
is $n_{H2}$ $\approx$ 2.8 $\times$ 10$^{5}$ cm$^{-3}$. Therefore, it is possible that the high-J CO
populations that drive the determination of T(CO) are significantly sub-thermal, which would
explain the temperature difference between the observed H$_2$ and CO populations. To explore this
possibility, we computed RADEX models~\citep{vandertak07} using the online
interface\footnote{http://var.sron.nl/radex/radex.php}.  For our measured values of T(H$_{2}$), 
N(CO), $b_{CO}$, and $n_{H}$ = 10$^{4}$ cm$^{3}$, we find that CO rotational levels 3~$\leq$~$J$~$\leq$~30
are characterized by temperatures less than the H$_{2}$ rotational temperature.  Finally, sub-thermal 
CO rotational temperatures are almost always observed in co-spatial CO and H$_{2}$ populations 
in the ISM \citep{burgh07,sheffer08}. We conclude that while there is uncertainty about the exact
spatial distribution of the gas and the viewing geometry, it is reasonable to infer that the H$_{2}$ 
and CO are co-spatial around HK Ori with CO/H$_{2}$ ratio $\approx 1.3 \times 10^{-5}$.   

The system radial velocity for HK Ori is $\approx 20-25$ km s$^{-1}$
\citep{reipurth96,baines04} which means the best-fit radial velocities for both the H$_2$
and CO absorption are blue-shifted from the system velocity by $\approx 35 - 40$ km s$^{-1}$
since we fit the spectrum in the Earth's rest frame. Again, it is worth noting
that the absolute velocity precision of COS is $\approx 15$ km s$^{-1}$. Even taking
into account the velocity precision of COS, the measured blue-shifts for the H$_2$
and CO absorption are large and inconsistent with a pencil beam observed through
a Keplerian disk since material in a circular disk along any line-of-sight should have velocities 
centered around $\approx 0$ km s$^{-1}$. However, it's possible that the absorption
arises at the base of a molecular disk wind launched magnetocentrifugally from the disk's upper 
atmosphere. Evidence of such a wind has been detected in CO emission from the Herbig Ae star 
HD 163296 by \citet{klaassen13} with characteristic velocities of $\approx 18$ km s$^{-1}$.
Thus depending on the wind launching angle and the disk inclination angle, blue-shifted
velocities of $\approx 40$ km s$^{-1}$ are not unreasonable if the absorption arises
in a disk wind.

Given the evidence for a disk or disk wind origin of the gas absorption towards
HK Ori, our derived CO/H$_2$ ratio of $\approx 1.3 \times 10^{-5}$ suggests significant
depletion of CO in the circumstellar material relative to the canonical molecular cloud 
value of $10^{-4}$ \citep{bergin17}. Lower values of CO/H$_2$ are expected for cooler 
disks with mid-plane temperatures $\lesssim 30-40$ K which is mainly driven by dust grain 
surface chemistry \citep{reboussin15}. However, the depletion factor of CO is highly 
dependent on age and location in the disk \citep[e.g.,][]{krijt20} which makes it difficult 
to say anything definitive about HK Ori's disk based on our derived ratio given the uncertainties 
around the line-of-sight probed by our observations. If our observations trace the
warm molecular layer of HK Ori's disk, the CO/H$_2$ ratio is line with the depleted
values reported for other protoplanetary disks \citep[e.g.,][]{favre13,mcclure16,schwarz16}.

\section{CONCLUSION}
\label{sec:conc}

We presented new FUV observations from the \textit{Hubble Space Telescope} 
of H$_2$ and CO absorption towards the Herbig Ae stars HK Ori and T Ori. We find
a significant column of CO absorption and a notable lack of H$_2$ absorption 
towards T Ori, which we posit is related to variable occultation
of the star by inner disk warping. HK Ori shows blue-shifted absorption in 
both H$_2$ and CO with derived temperatures and column densities which 
place the gas firmly in the circumstellar environment
around HK Ori. Although there is considerable uncertainty as to the exact
line-of-sight traced by our observations, the derived CO/H$_2$ value around
HK Ori is $\approx 1.3 \times 10^{-5}$ suggesting a depleted reservoir of CO.
Spatially-resolved molecular emission maps of both systems will help place our
pencil-beam observations in context and provide a more precise understanding
of the CO/H$_2$ value derived for the gas around HK Ori. 

\bigskip

{\bf Acknowledgments:} We thank the referee for their suggestions, which
helped improve the quality of the manuscript. P. W. C. acknowledges useful 
conversations with Nicole Arulanantham about the molecular components of 
inner disks. The data in this paper were obtained through
\textit{HST} Guest Observing programs 12996 and 15070. This work has made use of 
NASA's Astrophysics Data System and the SIMBAD database,
which is operated at CDS, Strasbourg, France.

\clearpage

\bibliography{references}{}
\bibliographystyle{aasjournal}

\end{document}